\documentclass[a4paper,12pt]{article}

\usepackage{jheppub}
\usepackage[dvipsnames]{xcolor} 
\usepackage{graphicx} 
\usepackage{latexsym}
\usepackage{amsmath,amsfonts,amssymb}
\usepackage{comment}
\usepackage{physics} 

\usepackage{tikz}
\usetikzlibrary{decorations.pathmorphing}	

\usepackage{tikz-feynman}
\tikzfeynmanset{compat=1.1.0} 

\usepackage{hyperref}
\hypersetup{
    colorlinks,
    linktoc=page,
    linkcolor=Cerulean,
    citecolor=NavyBlue,
    urlcolor=Cerulean 
}

\newcommand{\be}{\begin{equation}}
\newcommand{\ee}{\end{equation}}
\newcommand{\bes}{\begin{eqnarray}}
\newcommand{\ees}{\end{eqnarray}}
\newcommand{\bdm}{\begin{displaymath}}
\newcommand{\edm}{\end{displaymath}}
\newcommand{\ba}{\begin{array}}
\newcommand{\ea}{\end{array}}
\newcommand{\pa}[1]{\left(#1\right)}
\newcommand{\paq}[1]{\left[#1\right]}

\newcommand{\kk}{{\mathbf k}}
\newcommand{\xx}{{\mathbf x}}

\newcommand{\GG}{\mathbb{G}}
\newcommand{\WW}{\mathbb{W}}

\global\long\def\cev#1{\reflectbox{\ensuremath{\!\!\vec{\reflectbox{\ensuremath{#1}\,\,}}}}}%

\begin{document}

\title{Gravitational Polarizability of Schwarzschild Black Holes}

\author[a]{Gabriel Vidal}
\emailAdd{gabriel.vidal@unesp.br}

\author[a]{Gabriel M. Dantas}
\emailAdd{gabriel.m.dantas@unesp.br}

\author[a,b]{Riccardo Sturani}
\emailAdd{riccardo.sturani@unesp.br}

\author[a]{Gabriel Menezes\footnote{On leave of absence from the Departamento de F\'{i}sica, Universidade Federal Rural do Rio de Janeiro.}}
\emailAdd{gabriel.menezes10@unesp.br}

\affiliation[a]{Instituto de F\'\i sica Te\'orica, UNESP - Universidade Estadual Paulista, S\~ao Paulo 01140-070, SP, Brazil}
\affiliation[b]{ICTP South American Institute for Fundamental Research, S\~ao Paulo 01140-070, SP, Brazil}

\abstract{
  The linear response of a Schwarzschild black hole to an external quadrupolar perturbation is studied in analogy to a mechanical electrodynamical system, with the goal to describe the gravitational polarizability. Its causality properties imply dispersion relations that relate fluctuation and dissipative properties. We review and combine results obtained via the Regge-Wheeler equation on one side and a perturbative, worldline effective field theory description on the other, obtaining a consistent description of the
  dispersion relations for the gravitational polarizability of a Schwarzschild black hole. We find that the classical part of the 2-point correlation function of the black hole multipole depends on the boundary conditions of the space-time the black hole is immersed in, which is relevant for the
  dispersion relations considered.
}


\maketitle

\section{Introduction}

With the advent of gravitational wave (GW) astronomy, triggered by the GW observatories LIGO \cite{LIGOSCientific:2014pky}, Virgo \cite{VIRGO:2014yos}, and KAGRA \cite{KAGRA:2020tym}, and their detections \cite{KAGRA:2021vkt}, both fundamental and phenomenological aspects of General Relativity (GR) have regained interest and are subject of vigorous investigations. All of the $\order{100}$ signals detected so far are interpreted as the outcome of coalescences of binary compact objects, and in the overwhelming majority of cases, at least one black hole (BH) is involved. While the moderate loudness of the signals detected did not allow the determination (or lack thereof) of finite size effects associated with BHs, it is expected that the third generation of GW detectors ET \cite{Punturo:2010zz} and CE \cite{Reitze:2019iox}, and the space detector LISA \cite{2017arXiv170200786A} will observe ``golden'' events, i.e., sufficiently loud detections where such feature of fundamental gravity can be observed or ruled out.

It is known that BHs in GR have zero \emph{static} tidal Love numbers \cite{Damour:2009vw,Binnington:2009bb,Kol:2011vg,Gurlebeck:2015xpa},
i.e., they do not deform under a static external perturbation. They can, however, absorb \cite{Page:1976df}. While the latter property can be considered intuitive, the former has been a puzzle from the effective field theory point of view,
as it corresponds to a fine-tuning of the coefficient of a specific Wilson
operator \cite{Porto:2016zng}, which recently has been explained in terms of
a symmetry enhancement \cite{Hui:2020xxx,Hui:2021vcv,Charalambous:2022rre}.

Tidal deformation and absorption properties as linear reponses to an external field
are elegantly related to one another
via dispersion relations. For example, in the worldline effective field theory (EFT) setup, the BH dynamics is effectively encoded in a point particle action supplemented by higher derivative operators, whose Wilson coefficients, or rather, their real and imaginary parts, are mapped to tidal deformability and absorption parameters, respectively. This procedure requires the use of the well-known Schwinger-Keldysh (or \emph{in-in}) formalism \cite{Schwinger:1960qe,Keldysh:1964ud,Goldberger:2019sya,Jones:2023ugm}.

Given this framework, it is tantalizing to try to relate the real and imaginary parts of the BH response to external fields with relations \`a la Kramers-Kronig, i.e., using dispersion relations. Naively, however, the vanishing of the static tidal deformation \emph{and} the presence of absorption are incompatible with dispersion relations -- or more broadly -- with the fluctuation-dissipation theorem, which, loosely speaking, forbids the existence of fluctuations without dissipation and vice-versa.\footnote{See \cite{Kehagias:2024rtz} for derivation of the vanishing of non-linear static Love number too.}

In this work we explain how this issue arises and we frame it in light of the current understanding of the BH linear response function (also called the \emph{gravitational polarizability}), complementing previous EFT-based results \cite{Goldberger:2004jt,Goldberger:2005cd,Goldberger:2019sya}.

The paper is structured as follows. In Sec.~\ref{sec:setup} we introduce the EFT setup and the standard Regge-Wheeler approach to gravitational scattering to pave the way for the EFT computation of the BH polarizability, which is performed by matching the numerical solution of the Regge-Wheeler equation to an EFT computation as proposed in \cite{Goldberger:2005cd}.\footnote{See also \cite{Jones:2023ugm}, where the \emph{absorptive} cross section is related to the \emph{quadrupole} two-point function.} In Sec.~\ref{sec:computation} we interpret our results in terms of dispersion relations for the BH polarizability. We summarize our findings in Sec.~\ref{sec:conclusions}.


\subsection*{Conventions}

We use the following convention for the Fourier transform
\begin{equation}
    \tilde{f}(\omega) \equiv \int \dd{t} e^{+i\omega t} f(t), \quad f(t) \equiv \int \frac{\dd{\omega}}{2\pi} \, e^{-i\omega t} \tilde{f}(\omega),
\end{equation}
which is the same adopted, e.g., in \cite{Maggiore:2018sht}. We use natural units for Newton's constant $G$ and the speed of light $c$, such that $G=c=1$. Planck's mass is defined as $M_{\text{Pl}}^2 \equiv 1/32\pi G$. The following shorthand notation is used in the text:
\begin{equation}
    \int_{\omega} \equiv \int \frac{\dd{\omega}}{2\pi}. 
\end{equation}
When using multi-index notation according to which the index $L$ denotes
$i_1i_2\dots i_l$ made of $l$ spatial symmetric and trace-free indices, so that
for $\ell=\ell'=2$ one has
\begin{equation}
  \delta^{LL'} = \delta^{i_1i_2,j_1j_2} \equiv \frac{1}{2} \pa{
    \delta^{i_1j_1} \delta^{i_2j_2} + \delta^{i_1j_2} \delta^{i_2j_1}-\frac 23\delta^{i_1i_2}\delta^{j_1j_2}}\,.
\end{equation}
%

\section{The setup}
\label{sec:setup}

\subsection{The EFT description}

Consider an effective worldline theory for a spinless BH. We want to model the
effects of a dynamically induced multipole. The starting point is a multipolar
action for a Schwarzschild BH of mass $M$, endowed with \emph{unknown} internal
degrees of freedom collectively denoted by $X$, which are responsible e.g. for
dissipation and quasi-normal mode excitations.
We make the following minimal assumption about these degrees of freedom: (i) Their dynamics is ruled by a generic, local Lagrangian ${\cal L}[X]$, and (ii) they determine the induced multipoles. 

In this work we focus on the electric quadrupole $Q_{ij}[X]$, which couples linearly
to the traceless, electric part of the Riemann tensor $E_{ij}$ as per the BH
worldline action \cite{Goldberger:2004jt}
\begin{equation}
    S_{\text{WL}} \supset \int \dd{\tau} \left( -M + \mathcal{L}[X]+ \frac{1}{2} Q_{ij}[X] E^{ij} + \cdots \right)\,,
\end{equation}
where dots standa for higher order electric multipoles and magnetic multipoles.
This reproduces the monopole and quadrupole part of the standard multipolar action of a compact body (Latin indices are purely spatial indices). Following the idea proposed in \cite{Goldberger:2005cd,Goldberger:2019sya}, one can write the
effective action obtained by ``integrating out'' the $X$s, which can be heuristically written as
\begin{equation}
    S_{\text{eff}}= -M \int \dd{\tau} + \int \dd{\tau} \int_{0}^{\infty} \dd{t} \chi_{ij,kl}(\tau -t)E^{ij}(t) E^{kl}(\tau),
\end{equation}
where it has been assumed that the induced quadrupole response to the external
field is \emph{causal} and \emph{linear}:
\begin{equation}
    Q_{ij}(t) = \int_{-\infty}^{t} \dd{t'} \chi_{ij,kl}(t-t') E^{kl}(t'),
\end{equation}
and where we have introduced the time-dependent polarizability tensor
$\chi_{ij,kl}(t)$, whose tensor structure $\chi_{ij,kl} = \frac{1}{2} \left( \delta_{ik} \delta_{jl} + \delta_{il} \delta_{jk} - \frac{2}{3} \delta_{ij} \delta_{kl} \right) \chi$ can be factorized, as required by spherical symmetry. Moving to Fourier space and Taylor expanding the (electric, quadrupolar) polarizability, one can write:
\begin{equation}\label{eq:eft_E2om}
    S_{\text{eff}}^{(E)} \sim \int \frac{\dd{\omega}}{2\pi} \left[ \tilde\chi_0-i\tilde\chi_{1} M\omega - \tilde\chi_{2}(M\omega)^2 + \cdots \right] |\tilde E_{ij}(\omega)|^2,
\end{equation}
where the coefficients $\tilde\chi_n$ are real numbers. Strictly speaking, this effective action needs to be recast in the \emph{in-in} formalism, as it will be discussed later. The small parameter of expansion here is $M \omega$, i.e., this is an expansion in derivatives of the external gravitational field, so that, for example, $\tilde \chi_0$ in eq.~(\ref{eq:eft_E2om}) is the only term for an instantaneous static response -- defining the standard static tidal Love number -- while the coefficients $\chi_{n\geq 1}$ are its dynamical generalizations. We remark that dynamical Love numbers present EFT running, as discussed, for instance, in Ref.~\cite{Ivanov:2024sds}, and the two-loop beta function for all scalar tidal operators exhibit a universal conservative contribution arising from the post-Minkowskian expansion.

Coefficients with even ($\tilde\chi_{2n}$) and odd indices ($\tilde\chi_{2n+1}$)  embody tidal and dissipative properties of the BH, respectively. As mentioned before, one should employ the in-in formalism \cite{Schwinger:1960qe,Keldysh:1964ud}, in which a doubling of the degrees of freedom allows one to capture dissipation effects while working at the level of the Lagrangian; otherwise terms involving \emph{odd} powers of $\omega$ would be total derivative terms. For simplicity, we have written a schematic in-out action in eq. \eqref{eq:eft_E2om}. In turn, absorption effects can also be addressed via on-shell scattering amplitudes, see for instance Ref.~\cite{Aoude:2024jxd}.

\subsection{The GR description}

The scattering of gravitational waves in a Schwarzschild background is a classical problem in GR, whose solution within the scope of the black-hole perturbation theory methods of Teukolsky and Press has been subject to many seminal studies. See, for example,~\cite{Page:1976df,Handler:1980un}. We summarize a small part of these findings in this section.

By decomposing the GW perturbation $h_{\alpha\beta} \equiv \sum_{a=1}^{10}\sum_{\ell,m}R^a_{\omega \ell m}(r) \left[T_{\ell m}^a(\theta,\phi)\right]_{\alpha\beta} \, e^{-i\omega t}$
using the \emph{tensorial} spherical harmonics $T_{\ell m}^a$,
one obtains a Schrödinger-like equation for the scalar radial function $R_{\omega \ell}(r)$:
\begin{equation}\label{eq:teu}
  \left[  \dv[2]{}{r_*} + \omega^2 - V_\ell(r) \right]
  \left(rR_{\omega \ell m}(r)\right) = 0,
\end{equation}
where $r_{*} \equiv r + 2M \log\left( \frac{r}{2M}-1 \right)$, and $V_\ell(r)$ is an $\ell$-dependent potential, that for axial perturbations (Regge-Wheeler) takes\footnote{Under parity transformation, which
  in polar coordinates can be written as $r\to r$, $\theta\to \pi-\theta$, $\phi\to \phi+\pi$, out of the
  10 tensor harmonics pick a factor $(-1)^l$ and are said to be \emph{polar}, the 3 remaining pick a factor
  $(-1)^{l+1}$ and are said to be \emph{axial}, or odd.}
the explicit form \cite{Regge:1957td}
\be
V^{(RW)}_l(r)=\pa{1-\frac{2M}r}\paq{\frac{l(l+1)}{r^2}-\frac{6M}{r^3}}\,.
\ee
For the the polar perturbation (Zerilli) the form of the potential is
quantitatively different, but in both cases cases $V(r)\sim 1/r^2$ for large $r$.

The scattering/absorption process can be described in GR with the Regge-Wheeler and Zerilli equations,
see, for example, chapter 12 of the standard textbook \cite{maggioreII}.
We consider the scattering of an incoming \emph{plane}
GW with wave vector $\kk=\omega {\bf \hat z}$ by the central potential
generated by the Schwarzschild black hole into an outgoing spherical
wave with the \emph{same} polarization.
To simplify the computation, we consider as in \cite{Chrzanowski:1976jb}
a Cartesian component, for which one can apply the standard formula
plane wave decompposition, \ref{eq:app_plane} and obtain the elastic, absortive
and total scattering cross sections $\sigma_{e,a,t}$ in terms of the real inelasticity
parameter $\eta_\ell$ ($0\leq\eta_\ell\leq 1$) and real phase shift $\delta_\ell$:
\begin{subequations}
\label{eq:sigmas}
  \begin{align}
      \sigma_{\ell=2,\text{e}} &= \frac{\pi}{\omega^2} \sum_m |1-\eta_2 e^{2i\delta_2}|^2, \label{eq:sigma.e}\\
      \sigma_{\ell=2,\text{a}} &= \frac{\pi}{\omega^2} \sum_m \big( 1 - \eta_2^2 \big), \label{eq:sigma.a}\\
      \sigma_{\text{t}} &\equiv \sigma_{\text{e}} + \sigma_{\text{a}},
  \end{align}
\end{subequations}
where $\sigma_{\text{e,a,t}}$ are respectively the elastic, absorption and total cross
sections.

\begin{figure}
    \centering
    \includegraphics[width=.48\linewidth]{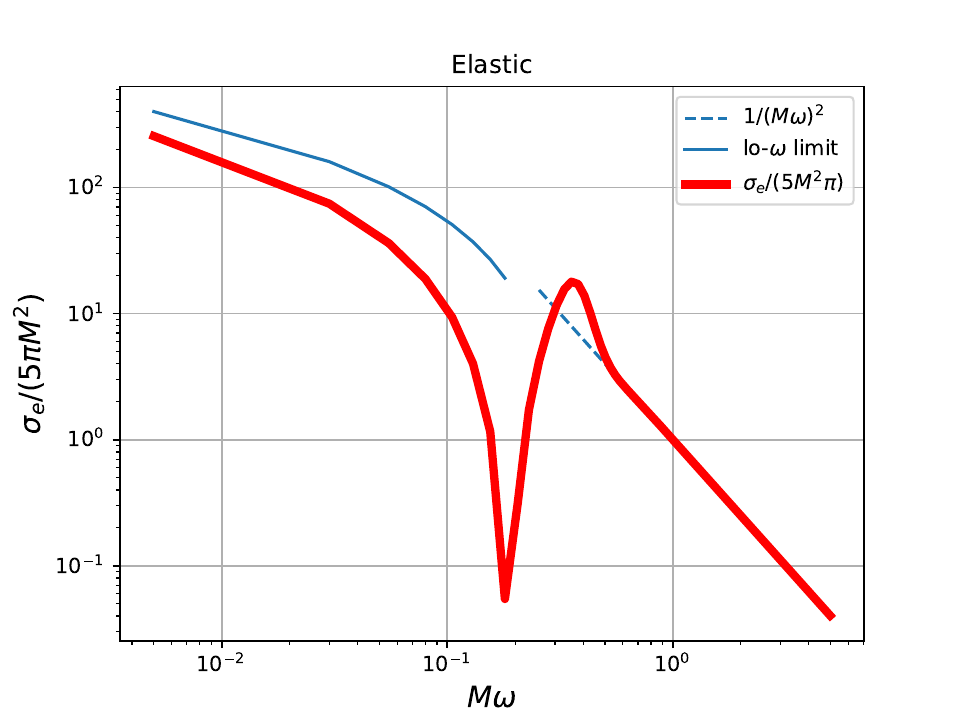}
    \includegraphics[width=.48\linewidth]{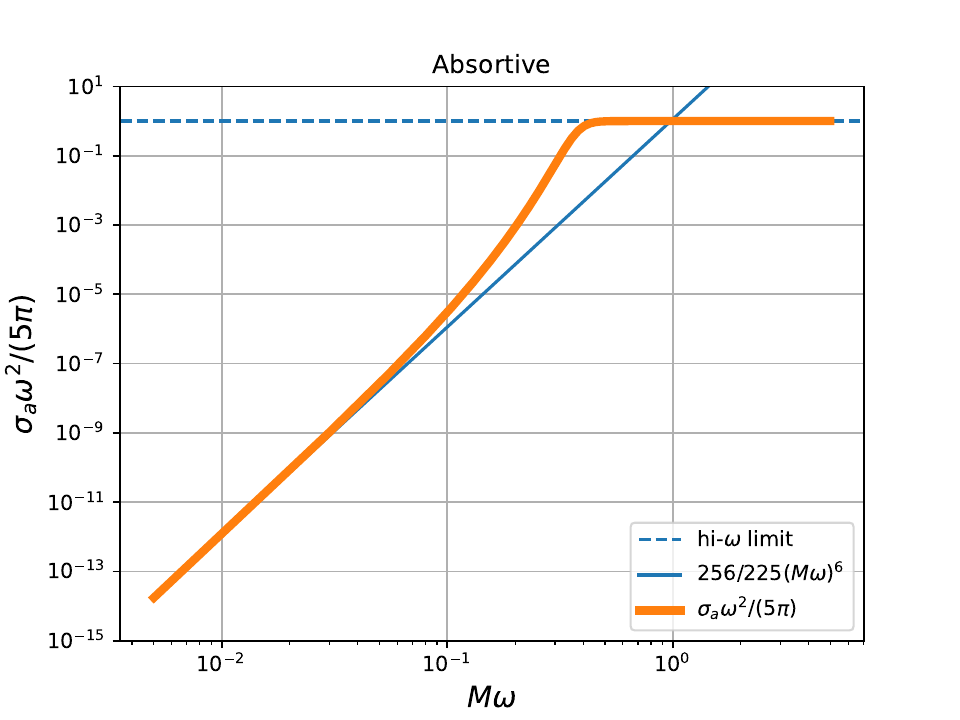}\\
    \caption{Normalized elastic (left) and absorption (right) cross sections and their
      asymptotic values.
      The $M\omega\to 0$ limit of $\sigma_{\text{e}}$ is given in (\ref{eq:sigmae}), whereas $\frac{\omega^2}{5\pi}\sigma_{\text{a},\ell=2}\to \frac{256}{225} (M\omega)^6$.
      For $M\omega\gg 1$, $\sigma_{\text{e,a},\ell=2}\to \frac{5\pi}{\omega^2}$.}
    \label{fig:sigmas}
\end{figure}

The solutions to \eqref{eq:teu} in the free case ($M=0$) come in two classes: pure left-moving $\cev{R}_{\omega \ell}(r)$ and right-moving $\vec{R}_{\omega \ell}(r)$ modes. Apart from an overall normalization factor, their asymptotic forms can be
parameterized as \cite{DeWitt:1975ys}
\begin{subequations}
  \label{eq:RAB}
  \begin{align}
        \cev{R}_{\omega \ell}(r) &= \frac{1}{r} 
          \begin{cases}
               B_{\ell}(\omega) \, e^{-i\omega r_{*}}, \quad & r \to 2M \\
               e^{-i\omega r_{*}} + \cev{A}_{\ell}(\omega) \, e^{i\omega r_{*}}, & r \to \infty
    \end{cases} \\[3pt]
    \vec{R}_{\omega \ell}(r) &= \frac{1}{r} 
          \begin{cases}
               e^{-i\omega r_{*}} + \vec{A}_{\ell}(\omega) \, e^{i\omega r_{*}}, \quad & r \to 2M \\
               B_{\ell}(\omega) \, e^{-i\omega r_{*}}, & r \to \infty
    \end{cases}
    \end{align}
\end{subequations}
The fact that $B_{\ell}$ is the same coefficient in both the left-moving and the right-moving mode is a simple consequence of the constancy of the Wronskian: 
\begin{equation}
    \vec{R}_{\omega \ell}(r) \, \frac{\dd \cev{R}_{\omega \ell}(r)}{\dd r_*} -\frac{\dd \vec{R}_{\omega \ell}(r)}{\dd r_*} \, \cev{R}_{\omega \ell}(r) = \text{constant in $r$},
\end{equation}
which also requires
\begin{equation}
    \vec{A}_{\ell} = - \frac{B_\ell^*}{B_\ell} \, \cev{A}_{\ell}.
\end{equation}
Moreover, probability conservation enforces $|A_\ell|^2+|B_\ell|^2=1$ (Notice that $|\vec A_\ell| = |\cev A_\ell|$). From our numerical solutions of the Regge-Wheeler equation \eqref{eq:teu} we obtain $\cev{R}_{\omega \ell}$, from which $\cev{A}_{\ell}(\omega)$ and $B_{\ell}(\omega)$ have been extracted. Their relation to $\sigma_{\text{e,a}}$ can be obtained by expanding the incident plane wave as in eq.~\eqref{eq:app_plane} and then following the standard textbook procedure, which gives
\begin{subequations}\label{eq:csab}
    \begin{align}
      \sigma_{\ell=2,\text{e}} &= \frac{\pi}{\omega^2} \sum_{m} |1 + \cev{A}_2
      |^2, \\[3pt]
        \sigma_{\ell=2,\text{a}} &= \frac{\pi}{\omega^2} \sum_{m}|B_{2}|^2.
    \end{align}
\end{subequations}
Our numerical results for $(\omega^2/5\pi) \, \sigma^{\text{e,a}}_{\ell=2}$ as a function
of $M\omega$ are reported in fig.~\ref{fig:sigmas}, where the limits $M\omega\to 0$ and $M\omega\gg 1$ are also shown. The $M\omega\to 0$ limit of $\sigma_{\text{a}}$ can be read from \cite{Page:1976df}:
\begin{equation}\label{eq:spage}
  \sigma^{\text{a}}_{\ell=2} = \frac{256\pi}{45} M^6 \omega^4.
\end{equation}
To compare it with the 
which plugged into \eqref{eq:sigma.a} gives
\begin{equation}\label{eq:eta2}
    \eta_2 \simeq 1 - \frac{128}{225} (M\omega)^6, \quad M\omega \ll 1,
\end{equation}
or $|B_{\ell}| \simeq \frac{16}{15} \,(M\omega)^3$ from \eqref{eq:csab} for $M\omega\ll 1$, which agrees with the value than can be deduced from eqs. (12) and (13) of \cite{Bautista:2024emt}.

The perturbative analytic result for the scattering phase in the Regge-Wheeler equation
is \cite{Poisson:1994yf,Bautista:2024agp}
\begin{equation} 
  \delta_2^{(\text{RW})} = \left( \log(4M\omega) + \gamma_{\text{E}}-\frac{5}{3} \right) 2M\omega + \frac{107\pi}{105}(2M\omega)^2+\order{(M\omega)^3},
\end{equation}
which together with \eqref{eq:eta2} gives
\begin{equation}\label{eq:sigmae}
    \begin{split}
        \sigma^{\text{e}}_{\ell=2} = 80\pi M^2 \bigg( \log(4M\omega) + \gamma_{\text{E}} - \frac{5}{3} \bigg) \bigg[ &\log(4M\omega) + \gamma_{\text{E}} - \frac{5}{3} \\[3pt]
        & + \frac{107\pi}{210} (2M\omega) + \order{(M\omega)^2} \bigg].
    \end{split}
\end{equation}

For $M\omega\gg 1$ one has $\eta_{\ell} \to 0$, leading to $A_\ell\to 0$, $B_\ell\to 1$, and $\sigma^{\text{e,a}}_{\ell=2}\to 5\pi/\omega^2$.

\section{The matching}
\label{sec:matching}

\subsection{Absorption cross section and quadrupole EFT two-point function}

While the scattering of radiation by longitudinal gravitational modes does not
give rise to absorption but is responsible for the leading order elastic scattering,
following \cite{Goldberger:2005cd} the leading order absorption process can be related
to the \emph{Feynman} two-point function:
\begin{equation}
  \label{eq:GF}
  iG_{\text{F}}^{LL'}(\omega)\equiv\int\frac{\dd t}{2\pi} \, e^{-i\omega t}\langle T\{Q^L(t)Q^{L'}(0)\}\rangle
\end{equation}
where $L,L'$ are collective indices, used here for the spin-2 representation of the rotation groups. Now we consider the coupling of $h_L$ to the source quadrupole $Q_L$: $\frac 1{4M_{\text{Pl}}}\ddot h_LQ_L$. A straightforward application of the optical theorem \cite{Goldberger:2005cd} for $\ell=2$ (in specific computations only the case $\ell=2$ will be considered) yields
\begin{equation}\label{eq:fey_opt}
  \sigma^{\text{a}}_{\ell=2} = 2\times \frac{16\pi}{2\omega} \, \omega^4
  \Im \sum_{\text{pols}} \tilde G_\text{F}^{LL'}(\omega)\epsilon_L\epsilon^*_{L'}.
\end{equation}
Therefore \footnote{To reinsert the powers of the Newton's constant, one can
  observe that (\ref{eq:GF}) has dimensions $[M]^2[L]^3$, hence a factor $G^4$ is
  required on the right hand side of (\ref{eq:IGF}).}, 
\begin{equation}
  \label{eq:IGF}
  \Im \tilde G_{\text{F}}^{LL'}(\omega) = \delta^{LL'}\frac{32}{45}M^6|\omega|,
\end{equation}
where the low energy value of $\sigma^{\text{a}}_{\ell=2}$ in eq. \eqref{eq:spage} has been used.

The \emph{Feynman} two-point function is related to the advanced and retarded ones via
\begin{equation}\label{eq:fey_ret}
    G_{\text{F}} = \frac{1}{2} \pa{G_{\text{A}} + G_{\text{R}}} - \frac{i}{2}\pa{\Delta_{+} + \Delta_{-}}= G_{\text{R}} - i\Delta_{-},
\end{equation}
where $\Delta^{L L'}_\pm(t)$ are the standard Wightman functions\footnote{Our $\Delta_{\pm}(\omega)$ are the $A_{\pm}(\omega)$ functions of \cite{Goldberger:2019sya}. We use $A$ to denote the reflection coefficients of eq. \eqref{eq:RAB}.} $\Delta_\pm^{LL'}$
\begin{subequations}\label{eq:Wight_pm}
    \begin{align}
        \Delta_{+}^{LL'}(t,t') &\equiv \langle Q^{L}(t) Q^{L'}(t') \rangle, \label{eq:Wight_plus}\\
        \Delta_{-}^{LL'}(t,t') &\equiv \langle Q^{L}(t) Q^{L'}(t') \rangle = \Delta_{+}^{L'L}(t',t), \label{eq:Wight_minus}
    \end{align}
\end{subequations}
and from now on we use rotational symmetry to impose $\Delta_\pm^{LL'}(t,t') =
\delta^{LL'}\Delta_\pm(t,t')$, so that
\begin{equation}
\label{eq:spherical}
    \tilde{\Delta}_{+}(\omega)=\tilde{\Delta}_{-}(-\omega).
\end{equation}
If one further assumes that $Q_L$ is an Hermitian operator ($Q_L=Q_L^\dagger$), one has
\begin{equation}
    \begin{split}
      \Tilde{\Delta}_{+}^{LL'}(\omega) = \int_{\omega} e^{i\omega(t-t')} \langle
      Q_L^\dagger(t) Q^\dagger_{L'}(t') \rangle
      &= \bigg( \int_{\omega} e^{-i\omega(t-t')}
      \langle Q_{L'}(t') Q_L(t)\rangle \bigg)^{*},
    \end{split}
\end{equation}
so that 
\begin{equation}\label{eq:Wight_star}
   \Tilde{\Delta}_{+}^{LL'}(\omega) = \big[ \tilde{\Delta}_{-}^{LL'}(-\omega) \big]^{*} = \big[ \tilde{\Delta}_{+}^{L'L}(\omega) \big]^{*} = \delta^{LL'} \tilde{\Delta}_{+}^{*}(\omega),
\end{equation}
in agreement with eq.~(3.24) of \cite{Goldberger:2020fot} and the assumption of spherical symmetry. See also App.~\ref{app:g2pw} for notation and details. 

\subsection{The gravitational Wightman function in the EFT setup}

Following the approach of \cite{Goldberger:2019sya}, we compute (perturbatively) the semiclassical graviton two-point function with EFT methods and compare it with the results coming from quantum field theory over a Schwarzschild background, which we refer to as the ``full theory.''

The presence of the BH modifies the Wightman function of the bulk
field $\WW_h^{LL'}\equiv \langle h^L(t,\vb{x}) h^{L'}(t',\vb{x}')\rangle_{\text{BH}}$, and these corrections can be computed perturbatively within the \emph{in-in} formalism in an EFT setup.
A short overlook of this formalism, as well as the details of our computation, are collected in the appendix \ref{app:g2pw} (see eq. \eqref{eq:2ptinin} in particular).
Here we focus on one specific (radiative) polarization of the gravitational perturbation $h$, obtaining the result the result which is accurate to leading-order
interactions with the BH, is\footnote{Crucially, we obtain a result that differs from the one reported in eq. (2.21) of \cite{Goldberger:2019sya}: the sign in front of their $\theta$ function should be changed to obtain our result. This small modification has important consequences for the matching procedure, as we shall discuss.} 

\begin{equation}
\label{eq:bulk2pt}
     \begin{split}
         \mathbb{W}_h^{LL'}(x,x') = \langle h^L(x) &h^{L'}(x') \rangle_{\text{M}} \\[3pt]
         &+ \frac{\delta^{LL'}}{(4\pi r)^2} \int_{\omega}  e^{-i\omega(t-t')} \big[ \theta(-\omega) \Delta_+(\omega) + \theta(\omega) \Delta_-(\omega) \big],
     \end{split}
\end{equation}
where by $\langle h^L(x) h^{L'}(x') \rangle_{\text{M}}$ we denote the two-point function in Minkowski space, i.e. without the BH contribution.
The second term in \eqref{eq:bulk2pt} is evaluated at the spatial coincidence limit $\vb{x}=\vb{x}'$, with $r=|\vb{x}|$.
This bulk EFT two-point function can be compared to the full theory one, which
can be found in literaturery with different boundary conditions,
the most common ones being the \emph{Boulware} (B), the \emph{Unruh} (U), and the \emph{Hartle-Hawking} (HH) states. 

From the perspective of quantum field theory over curved spacetimes, such conditions reflect the non-uniqueness of the vacua when Poincaré symmetry is absent \cite{Birrell:1982ix}. The states mentioned above correspond, respectively, to the ``no radiation'', the ``only outgoing radiation'', and the ``thermal bath'' boundary conditions. Adapting the classical result of \cite{Candelas:1980zt}, the explicit form of our bulk field two-point function (\ref{eq:bulk2pt}) can be
matched to the full thoery ones, which we compute here below for the three
aforementioned boundary conditions. 

To ease the matching, it is convenient to choose spatial points with same
angular variables:
${\bf x}=r{\bf \hat n}$, ${\bf x'}=r'{\bf \hat n}$.
This allows to use
the coincidence limit of the addition theorem for tensorial
spherical harmonics \cite{Michel2019,Monteverdi:2024xyp},
projected onto the radiative polarization $\epsilon$
\be
\epsilon_L\epsilon_{L'}^*\sum_m[T^a_{\ell m}(\theta,\phi)]_L[T^b_{\ell m}(\theta,\phi)]_{L'}=
\frac{2\ell+1}{4\pi}\delta^{ab}|\epsilon|^2\,.
\ee

\subsubsection{Boulware}
In the Boulware case one has:
\begin{equation}
    \begin{split}
        \epsilon_L\epsilon_{L'}^*\langle h^L(t,\vb{x}) h^{L'}(t',\vb{x}') \rangle_{\text{B}}= |\epsilon|^2 &\int_{0}^{\infty} \frac{\dd{\omega}}{(4\pi)^2\omega} \, e^{-i\omega(t-t')}\sum_\ell (2\ell+1) \\
        &\times \big[ \vec{R}_{\omega \ell}(r) \vec{R}^{*}_{\omega \ell}(r') + \cev{R}_{\omega \ell}(r) \cev{R}^{*}_{\omega \ell}(r') \big]\,.
    \end{split}
\end{equation}
Expanding this result for large spatial coordinates eq.~(\ref{eq:RAB}), one obtains
\begin{equation}\label{eq:hhB}
    \begin{split}
      \epsilon_L\epsilon_{L'}^*\langle h^L(t,\vb{x}) &h^{L'}(t',\vb{x}') \rangle_{\text{B}} \simeq |\epsilon|^2 \int_{0}^{\infty} \frac{\dd{\omega}}{16\pi^2 \omega} \frac{e^{-i\omega(t-t')}}{rr'} \sum_\ell (2\ell+1) \\
        &\times \bigg[ |B_{\ell}|^2 e^{i\omega(r_{*}-r_{*}')} + \big( e^{-i\omega r_{*}} + \cev{A}_{\ell} \, e^{i\omega r_{*}} \big) \big( e^{i\omega r_{*}'} + \cev{A}^{*}_{\ell} e^{-i\omega r'_{*}} \big) \bigg]\,.
    \end{split}
\end{equation}

To ease the physical interpretation of the above expression let us compare it with the corresponding Minkowski two-point function \eqref{eq:Wmink}, expanded in the same limit: 
\begin{equation}
    \begin{split}
        \epsilon_L\epsilon_{L'}^*\langle h^L(t,\vb{x}) h^{L'}(t',\vb{x}') \rangle_{\text{M}} &\simeq \frac{|\epsilon|^2}{rr'} \int_{0}^{\infty} \frac{\dd{\omega}}{16\pi^2 \omega} \, e^{-i\omega(t-t')} \sum_\ell\, (2\ell+1) \\ 
        & \quad \quad \times \big(e^{-i\omega r} - (-1)^{\ell} e^{i\omega r} \big) \big( e^{i\omega r'} - (-1)^{\ell} e^{-i\omega r'} \big). 
    \end{split}
\end{equation}

By straightforward algebraic manipulations, \eqref{eq:hhB} can be recast into
\begin{equation}\label{eq:hhB2}
    \begin{split}
       \epsilon_L\epsilon^*_{L'}\langle h^L(t,r\vu{n}), \,h^{L'}(t',r\vu{n}') \rangle_{\text{B}} &\simeq \epsilon_L\epsilon^*_{L'}\langle h^L(t,r\vu{n}), h^{L'}(t',r\vu{n}') \rangle_{\text{M}} \\[3pt]
        &+ \frac{|\epsilon|^2}{r^2} \int_{0}^{\infty} \frac{\dd{\omega}}{16\pi^2 \omega} \, e^{-i\omega(t-t')} \sum_{\ell=0}^{\infty} \, (2\ell+1) P_{\ell}(\cos \gamma) \\ 
        &\times \big[ |B_{\ell}|^2 + |\cev{A}_{\ell} + (-1)^{\ell}|^2 - 2 \big( 1 + (-1)^{\ell} \Re \cev{A}_{\ell} \big) \big]\,,
    \end{split}
\end{equation}
where fast oscillating exponentials $e^{\pm i\omega(r'+r)}$ have been dropped, in light of the large $r,r'$ limit and taken the limit $r=r'$.
The last line of \eqref{eq:hhB2} vanishes by virtue of $|\cev{A}_{\ell}|^2+|B_{\ell}|^2=1$. 

Using the symmetry relation \eqref{eq:Wight_minus}, the matching of \eqref{eq:hhB2} onto \eqref{eq:bulk2pt} leads to the condition $\Delta_{+}(\omega) \propto \theta(\omega)$, meaning that the Wightman function $\Delta_{+}$ has no support over negative frequencies.
Even though the same conclusion was reached in reference \cite{Goldberger:2019sya}, our EFT computation \eqref{eq:bulk2pt} is different, and so is the matching computation.
In \cite{Goldberger:2019sya}, the following asymptotic mode summation is used:
\begin{equation}
  \label{eq:RleftM}
    \sum_{\ell=0}^{\infty} \, (2\ell+1)|\cev{R}_{\ell}(\omega|r)|^2 \sim 4\omega^2, \quad r\to \infty\,,
\end{equation}
which disregards all non-Minkowski terms, i.e. terms involving $\cev A_\ell$, see eq.~(\ref{eq:RAB}); in particular it does not capture
all the contributions involving $\Delta_{+}(\omega)$, the $A_+$ of
\cite{Goldberger:2019sya}.\footnote{Using eq.~(\ref{eq:RleftM}),
  the authors found that the Boulware matching yields a non-vanishing
  contribution to the bulk two-point function from the interaction with the BH,
  implying that (to leading order in $M\omega$ and $r=r' \to \infty$)
  $\mathbb{W}_h^{LL'}(x,x') = \langle h^L(x) h^{L'}(x') \rangle_{\text{M}} + \frac{1}{4\pi^2} \pa{\frac{2M}{r}}^2 \int_{0}^{\infty} \dd{\omega} \omega \, e^{-i\omega(t-t')}$.}

As a consistency check of our results we verify that the radial flux of radiation also vanishes, as expected in the Boulware state:
\begin{equation}
    \begin{split}
        \langle \text{in}| T^{rt}(x) |\text{in}\rangle &= \frac{1}{2} \lim_{x \to x'} (\partial_r \partial_{t'} + \partial_t \partial_{r'}) \mathbb{W}_h(x,x') \\[5pt]
        &= \frac{1}{16\pi^2 r^2} \int_{\omega} \omega^2 e^{-i\omega(t-t')} \big[ \theta(-\omega) \Delta_{+}(\omega) + \theta(\omega) \Delta_{-}(\omega) \big] \\[5pt]  
        &= 0\,.
    \end{split}
\end{equation}

\subsubsection{Unruh} Next, we consider the matching onto the Unruh state. As before, the starting point is the result from quantum field theory over a Schwarzschild background: 
\begin{equation}
    \begin{split}
      \epsilon_L\epsilon_{L'}^*\langle h^L(t,r\vb{n}) h^{L'}(t',r'\vb{n}) \rangle_{\text{U}} = |\epsilon|^2 &\int_{-\infty}^{\infty} \frac{\dd{\omega}}{(4\pi)^2\omega} \, e^{-i\omega(t-t')} \sum_\ell(2\ell+1)\\
        &\times \bigg[ \frac{\vec{R}_{\omega \ell}(r) \vec{R}^{*}_{\omega \ell}(r')}{1-e^{-8\pi M\omega}} + \theta(\omega) \cev{R}_{\omega \ell}(r) \cev{R}^{*}_{\omega \ell}(r') \bigg]\,,
    \end{split} 
\end{equation}
whose large $r,r'$ limit yields
\begin{equation}
    \begin{split}
      \epsilon_L\epsilon_{L'}^* \langle h_{L}(t,\vb{x}) h_{L'}(t',\vb{x}') \rangle_{\text{U}} \simeq |\epsilon|^2 \int_{-\infty}^{\infty} &\frac{\dd{\omega}}{16\pi^2 \omega} \frac{e^{-i\omega(t-t')}}{rr'} \sum_\ell \, (2\ell+1)  \\
       &\times \bigg( \frac{|B_{\ell}|^2}{1-e^{-8\pi M\omega}} \, e^{i\omega(r_{*}-r'_{*})} + \theta(\omega) \mathcal{U}_{\omega \ell}(r,r') \bigg), 
    \end{split}
\end{equation}
where we have defined $\mathcal{U}_{\omega \ell}(r,r')$ as 
\begin{equation}
    \begin{split}
        \mathcal{U}_{\omega \ell}(r,r') &\equiv \big[e^{-i\omega r} - (-1)^{\ell} e^{i\omega r} \big] \big[ e^{i\omega r'} - (-1)^{\ell} e^{-i\omega r'} \big] \\[5pt]
        &+ |\cev{A}_{\ell} + (-1)^{\ell}|^2 e^{i\omega(r-r')} - 2 \big[1 + (-1)^{\ell} \Re \cev{A}_{\ell} \big] e^{i\omega(r-r')} \\[5pt]
        &+ \big[ \cev{A}_{\ell}^{*} + (-1)^{\ell} \big] e^{-i\omega(r+r')} + \big[ \cev{A}_{\ell} + (-1)^{\ell} \big] e^{i\omega(r+r')}. 
    \end{split}
\end{equation}
Specializing to $r=r' \to \infty$, we find (again after dropping fast oscillating terms )
\begin{equation}\label{eq:2ptUM}
    \begin{split}
      \epsilon_L\epsilon_{L'}^*\langle h^L(t,r\vu{n}) &h^{L'}(t',r\vu{n}) \rangle_{\text{U}} = \epsilon_L\epsilon_{L'}^*\langle h^L(t,r\vu{n}) h^{L'}(t',r\vu{n}) \rangle_{\text{M}}  \\[3pt]
        &+ \frac{|\epsilon|^2}{r^2} \int_{-\infty}^{\infty} \frac{\dd{\omega}}{16\pi^2 \omega} \, e^{-i\omega(t-t')} \sum_{\ell=0}^{\infty} \, (2\ell+1) \, \times\\[3pt]
        &\quad\times\frac{|B_{\ell}|^2}{1-e^{-8\pi M\omega}} + \theta(\omega) \big\{ |\cev{A}_{\ell} + (-1)^{\ell}|^2 - 2\big[1 + (-1)^{\ell} \Re \cev{A}_{\ell} \big] \big\}\,.
        \end{split}
\end{equation}
Remarkably, the $1/r^2$ part of (\ref{eq:2ptUM}) can be re-arranged as
\begin{equation}
    \frac{|B_{\ell}|^2}{e^{8\pi M|\omega|}-1} \big[ \theta(\omega)-\theta(-\omega) \big] = \frac{|B_{\ell}|^2}{e^{8\pi M|\omega|}-1} \, \textsf{sgn}(\omega),
\end{equation}
where $\textsf{sgn}$ is the sign function. Our final expression for the full theory result is thus
\begin{equation}\label{eq:2ptUMbis}
    \begin{split}
        &\epsilon_L\epsilon_{L'}\langle h^L(t,r\vu{n}) h^{L'}(t',r\vu{n}) \rangle_{\text{U}} = \epsilon_L\epsilon_{L'}\langle h^L(t,r\vu{n}) h^{L'}(t',r\vu{n}) \rangle_{\text{M}}  \\[3pt]
        &+\frac{|\epsilon|^2}{r^2} \int_{-\infty}^{\infty} \frac{\dd{\omega}}{16\pi^2 \omega} \, e^{-i\omega(t-t')} \sum_\ell \, (2\ell+1)\, \frac{|B_{\ell}|^2}{e^{8\pi M|\omega|}-1} \, \textsf{sgn}(\omega).
    \end{split}
\end{equation}
Matching to the perturbative computation (\ref{eq:bulk2pt}) one gets:
\begin{subequations}
  \label{eq:DnegU}
  \begin{align}
        \Delta_{-}(\omega > 0) &= \frac{5}{2\pi|\omega|} \frac{|B_2|^2}{e^{8\pi M|\omega|}-1},\\[3pt]
        \Delta_{+}(\omega < 0) &= \frac{5}{2\pi|\omega|} \frac{|B_2|^2}{e^{8\pi M|\omega|}-1}\,,
    \end{align}
\end{subequations}
which is consistent with the ``spherical symmetry condition''
\eqref{eq:spherical} and the hermiticity condition \eqref{eq:Wight_star}.

For the energy-momentum tensor in the perturbative computation one gets
\begin{equation}\label{eq:TtrU}
    T^{tr} = \frac{\omega^2}{16\pi^2r^2} \big[ \theta(\omega) \Delta_-(\omega) + \theta(-\omega) \Delta_+(\omega) \big] = \frac{4\omega}{e^{8\pi M |\omega|}-1} \, |B_{\ell}|^2,
\end{equation}
which is consistent with a net flux of particles escaping to $r\to \infty$.

\subsubsection{Hartle-Hawking} When the BH is in equilibrium with a thermal bath
, the full theory two-point function is 
\begin{equation}
    \begin{split}
      \epsilon_L\epsilon_{L'}\langle h^L(t,r\vb{n}) &h^{L'}(t',r'\vb{n}) \rangle_{\text{H}} = |\epsilon|^2\int_{-\infty}^{\infty} \frac{\dd{\omega}}{(4\pi)^2\omega}
      \sum_\ell (2\ell+1) \\[3pt]
        & \times\left[ e^{-i\omega(t-t')} \frac{\vec{R}_{\omega \ell m}(r) \vec{R}^{*}_{\omega \ell m}(r')}{1-e^{-8\pi M \omega}} + e^{i\omega(t-t')}\frac{\cev{R}^{*}_{\omega \ell m}(r) \cev{R}_{\omega \ell m}(r')}{e^{8\pi M \omega}-1} \right]\\
 &\stackrel{r=r'\to \infty}{=} \delta_{LL'} \int_{-\infty}^{\infty} \frac{\dd{\omega}}{16\pi^2\omega} \frac{e^{-i\omega(t-t')}}{rr'} \sum_{\ell}  (2\ell+1)\\
        & \times \frac 1{1-e^{-8\pi M \omega}}\paq{|B_\ell|^2+\mathcal{U}_{\omega \ell}(r,r')}\,.
    \end{split}
\end{equation}
Matching of the result above with the thermal ($\beta$) Minkowski two-point function
\begin{equation}
  \label{eq:hhHH}
    \langle h_L(t,r\vu{n}) h_{L'}(t',r\vu{n}) \rangle^{\beta}_{\text{M}} \equiv \frac{\delta_{LL'}}{2\pi} \int_{-\infty}^\infty \frac{\dd{\omega}}{2\pi} \frac{\omega}{1-e^{-8\pi M \omega}}\,,
\end{equation}
implies that 
\begin{equation}
    \theta(-\omega) \Delta_{+}(\omega) + \theta(\omega) \Delta_{-}(\omega) = 0,
\end{equation}
which is the same result of the Boulware state computation. Analogously, the flux at infinity derived from the energy momentum tensor vanishes, leaving only the flat space thermal contribution. This is consistent with the results presented in table I of \cite{Candelas:1980zt}.

\section{Gravitational dispersion relations}
\label{sec:computation}

We are now going to relate the real and imaginary part of the polarizability, which is the retarded two-point function of the quadrupole operator.

The retarded Green's function is real in direct space, hence
\begin{equation}
  \label{eq:GRstar}
  \tilde G_{\text{R}}^{LL'}(\omega)= \paq{\tilde G_{\text{R}}^{LL'}(-\omega)}^{*}\,,
\end{equation}
then using the dispersive representation of the retarded Green's function 
(\ref{eq:GRdisp}), the standard expression (real line integral case of the Sokhotski-Plemelj formula)
\begin{equation}
    \lim_{\epsilon \to 0} \frac{1}{x-i\epsilon} = \textsf{PV} \,  \frac{1}{x} + i\pi\delta(x)\,,
\end{equation}
the spherical symmetry relation \eqref{eq:spherical} and the reality condition \eqref{eq:Wight_star}, one obtains
($\Delta(\omega)\equiv\tilde \Delta_+(\omega)$)
\begin{subequations}\label{eq:GR_W}
    \begin{align}
      \Re\tilde G_{\text{R}}^{LL'}(\omega) &= \delta^{LL'}\textsf{PV}
        \int_{-\infty}^{\infty} \frac{\dd{\omega'}}{2\pi} \frac{\Delta(\omega') - \Delta(-\omega')}{\omega - \omega'} \\[3pt]
      \Im \tilde G_{\text{R}}^{LL'}(\omega) &= -\frac{\delta^{LL'}}{2}
        \paq{\Delta(\omega)-\Delta(-\omega)}\,,
    \end{align}
\end{subequations}

which is in agreement with equations (3.36) and (3.37) of \cite{Goldberger:2020fot}. Moreover, it can be verified that the large distance limit of the equal
time commutator $[h_L(t,r{\bf \hat n}),h_{L'}(t,r'{\bf\hat n'})]$ vanishes,
for both the EFT formula \eqref{eq:bulk2pt} and the result of the full theory
with the three different boundary conditions
\eqref{eq:hhB},\eqref{eq:2ptUMbis},\eqref{eq:hhHH}.

Using the analyticity property of $\tilde G_{\text{R}}^{LL'}$ (and that $\tilde G_R(\omega)\to 0$ as $\omega\to\infty$ in the complex plane) one gets
\begin{equation}\label{eq:gr_ri}
    \tilde G_{\text{R}}^{LL'}(\omega) = \frac{1}{2\pi i} \oint \frac{\dd{\omega'}}{2\pi} \frac{\tilde G_{\text{R}}^{LL'}(\omega')}{\omega' - \omega - i\epsilon} = \frac{1}{2\pi i} \, \textsf{PV} \int \frac{\dd{\omega'}}{2\pi} \frac{\tilde G_{\text{R}}^{LL'}(\omega')}{\omega' - \omega} + \frac{1}{2} \tilde G_{\text{R}}^{LL'}(\omega),
\end{equation}
from which the usual Kramers-Kronig relations follow
\begin{subequations}
    \begin{align}
        \Re \tilde G_{\text{R}}^{LL'}(\omega) &= \frac{1}{\pi} \, \textsf{PV} \int \frac{\dd{\omega'}}{2\pi} \frac{\Im \tilde G_{\text{R}}^{LL'}(\omega')}{\omega' - \omega}, \\[3pt]
        \Im \tilde G_{\text{R}}^{LL'}(\omega) &= -\frac{1}{\pi} \, \textsf{PV} \int \frac{\dd{\omega'}}{2\pi} \frac{\Re \tilde G_{\text{R}}^{LL'}(\omega')}{\omega' - \omega}\,.
    \end{align}
\end{subequations}
Using \eqref{eq:fey_opt} and \eqref{eq:fey_ret} one has
\begin{subequations}
\label{eq:KKr}
  \begin{align}
    \Re G_{\text{R}}^{LL'}(\omega) &= \Re G_{\text{F}}^{LL'}(\omega)\,, \\[3pt]
    \Im G_{\text{R}}^{LL'}(\omega) &= \Im G_{\text{F}}^{LL'}(\omega)+
    \delta^{LL'}\Delta(-\omega) \simeq \delta^{LL'} \left[ \frac{32}{45}M^6|\omega| + \Delta(-\omega) \right]\,,
  \end{align}
\end{subequations}
where for $\Im G_F$ we substituted its low $M\omega$ limit (\ref{eq:fey_opt}).

The vanishing of the static response $\tilde G_{\text{R}}(0) = 0$, leads to a sum rule \cite{Rothstein:2014sra} that can be rewritten as
\begin{equation}\label{eq:sum_rule}
    \int_{0}^{\infty} \dd{\omega} \frac{\Delta(\omega) - \Delta(-\omega)}{\omega} = 0\,.
\end{equation}
According to our results, in the {\bf B} and {\bf H} vacuum
(where $\Delta(\omega)\propto\theta(\omega)$), the above sum rule requires that
$\Delta(\omega>0)$ cannot be always positive, which is problematic as
$\sigma_{\text{A}}\propto \Delta(\omega)$.
On the other hand in the {\bf U} vacuum the sum rule \eqref{eq:sum_rule} can be
fulfilled non-trivially as $\Delta(-\omega) \neq 0$, see eq.~\eqref{eq:DnegU}.

This is in \emph{apparent} contrast to the common belief that retarded/advanced Green functions, constructed out of Wightman function commutator, embody the purely classical information, which being a c-number compared to quantum operators, cannot have vacuum-dependent expectation values. Indeed the ``quantum" piece of the Feynman Green's function is proportional to the symmetric combination of Wightman functions, see eq.~(\ref{eq:greendefs}c).
However the classical Green's function can depend on boundary conditions, which are implied by the choice of vacuum, see an analog result for the electromagnetic field in Minkowski and Schwarzschild space-times \cite{Zhu:2006wt,Zhou:2012eb}.


\section{Discussion}
\label{sec:conclusions}

We have analyzed the apparent puzzle of a black hole having vanishing static tidal deformability and absorptive properties, which is at odds with general principles based on causality and linear response. In particular we find that with Unruh boundary conditions, corresponding to a black hole emitting Hawking radiation, are compatible with the quadrupole two-point function having support for negative frequencies, as required by the net flux of radiation to infinity leaving, welcome property for the causality relations \`a la Kramers-Kronig.

We note however that by taking into account terms beyond the lowest order coupling to the quadrupole two point functions, one should include the tail process. 
As pointed out by \cite{DeLuca:2024ufn}, the tail effect, i.e., the scattering of radiation off the static curvature induced by BH, multiplies the response function by a complex number, which can have the effect of mixing up real and imaginary part, and change the holomorphic property of the retarded function because of the branch cut introduced by the logarithmic tail.

Actually, the effective field theory derivation presented in this work cannot catch a possible analytic contribution, which is not captured by imaginary part of the retarded Green's function, or the absorptive cross section. This is well known from methods based on quantum scattering amplitudes -- there are terms that can be reproduced by using unitarity cuts (which are in turn associated with discontinuities along branch cuts), but there are also contact terms that obviously do not contribute to those cuts. For instance, the use of the Cauchy's theorem to the Feynman propagator in quantum field theory makes use of a contour which contains contributions from arcs at infinity, and this may result (or not) in polynomial subtraction terms~\cite{Zwicky:2016lka}.

This is also known from Britto-Cachazo-Feng-Witten-type recursion relations, in which the shifted amplitude, as a function of the shift parameter $z$, can be written in terms of a pole term and a purely polynomial piece -- the constant term from this piece is what we call the residue of the pole at $z=\infty$, the boundary term.

This discussion implies that the real part of the retarded Green's function could in principle contain an undetermined real polynomial piece $F(\omega)$ \cite{Charalambous:2022rre} that should be added to the right hand side of (\ref{eq:KKr}a),
and which can can be written in terms of a polynomial in $\omega$. This polynomial piece corresponds to the Love numbers~\cite{Charalambous:2022rre}.

Now at $\omega \to 0$, due to the vanishing of static Love numbers for Schwarzschild black holes, the aforementioned sum rule \eqref{eq:sum_rule} would gain a constant terms outside the integral sign.
The physical interpretation of this sum rule may be understood by taking into account late-time tail effects as one can rewrite the dispersion-integral representation of the retarded Green's function in such a way that a connection between the near and wave zones is manifest~\cite{DeLuca:2024ufn}. At low frequencies the scattering is dominated by the leading instantaneous contribution associated with tidal effects, however, also at low frequencies the overlap between potential and radiation modes produces tail contributions, so physically the modified sum rule should be interpreted in terms of an overlap between tidal response and tail effects.


\section*{Acknowledgments}
The work of RS is supported by FAPESP grant n. 2022/06350-2 and 2021/14335-0,
and by CNPq n. 312320/2018-3. GMD is supported by FAPESP grant n. 2023/00295-2. GMe acknowledges partial support from CNPq under grant 102305/2024-2, FAPESP under grant n. 2023/06508-8 and FAPERJ under grant E-26/201.142/2022.


\appendix

\section{Useful equations}\label{app:uf}

A scalar plane wave with wave-vector $\kk=\omega{\bf \hat z}$ can be decomposed into radial ones via
\begin{equation}
  \label{eq:app_plane}
    e^ {i\kk \cdot {\bf x}}  = \frac{1}{2i\omega r} \sum_{\ell = 0}^{\infty} \, (2\ell+1) P_{\ell}(\cos\theta) \big[ (-1)^{\ell+1} e^{-i\omega r} + e^{i\omega r} \big],
\end{equation}
where $r\equiv |\xx|$, and $\theta$ is the polar angle of $\xx$ with the ${\bf \hat z}$ direction. The asymptotic solution of eq.~\eqref{eq:teu} when the leading term of $V(r)$ at infinity is the centrifugal barrier $\ell(\ell+1)/r^2$ is
\begin{equation}\label{eq:app_asym}
    R_{\ell} \sim \frac{1}{ir} \big[ (-i)^{\ell} e^{i(\omega r + \delta_{\ell})} - i^{\ell} e^{-i(\omega r + \delta_{\ell})} \big],
\end{equation}
where $\delta_{\ell}$ is complex in case of absorption and real in case of elastic scattering.

\section{Green and two-point functions}\label{app:g2pw}
We report below some generic formulae relating two-point functions to various type of Green functions. We denote the standard two-point Wightman function $W$ of a generic bulk field $\Phi^L$ as
\begin{subequations}
    \begin{align}
        \WW^{LL'}_+(x-x') \equiv \langle \Phi^{L}(x) \Phi^{L'}(x') \rangle, \\[3pt]
        \WW^{LL'}_-(x-x') \equiv \langle \Phi^{L'}(x') \Phi^{L}(x) \rangle,
    \end{align}
\end{subequations}
where the multi-index notation $L\equiv i_1 i_2 \cdots i_\ell$ is used. Then, the retarded, advanced, Feynman and Dyson Green's functions can be expressed respectively as
\begin{subequations}\label{eq:Greens}
    \begin{align}
        \GG^{LL'}_{\text{R}}(t,\vb{x}) &= -i \theta(t)  \big[ \WW_{+}^{LL'}(t,\vb{x}) - \WW_{-}^{LL'}(t,\vb{x}) \big], \\[3pt]
        \GG^{LL'}_{\text{A}}(t,\vb{x}) &=  i \theta(-t) \big[ \WW_{+}^{LL'}(t,\vb{x}) - \WW_{-}^{LL'}(t,\vb{x}) \big],\\[3pt]
        \GG^{LL'}_{\text{F}}(t,\vb{x}) &= -i \big[ \theta(t) \WW_{+}^{LL'}(t,\vb{x}) + \theta(-t) \WW_{-}^{LL'}(t,\vb{x}) \big], \\[3pt]
        \GG^{LL'}_{\text{D}}(t,\vb{x}) &= -i \big[ \theta(t) \WW^{LL'}_{-}(t,\vb{x}) + \theta(-t) \WW_{+}^{LL'}(t,\vb{x}) \big],
    \end{align}
\end{subequations}
in terms of the Wightman functions $\WW_{\pm}^{LL'}(x-x')$, which have the symmetries
\begin{equation}\label{eq:W_pm.syms}
  \WW_{\mp}^{L'\!L}(-t,-\vb{x}) =\WW_{\pm}^{LL'}(t,\vb{x})=\big[\WW_{\mp}^{LL'}(t,\vb{x})\big]^{*},
\end{equation}
for any Hermitian operator $\Phi$. From the property \eqref{eq:W_pm.syms} one can deduce the relations
\begin{equation}\label{eq:GD.GA.from.GF.GR}
  \begin{aligned}
    \GG_{\text{D}}^{LL'}(t,\vb{x}) &= - \big[ \GG_{\text{F}}^{LL'}(t,\vb{x}) \big]^{*}, \\[3pt]
    \GG_{\text{A}}^{LL'}(t,\vb{x}) &= \GG_{\text{R}}^{L'L}(-t,-\xx).
  \end{aligned}
\end{equation}
Using the identity $\theta(t)+\theta(-t)=1$, the definitions (\ref{eq:Greens}), and the properties (\ref{eq:GD.GA.from.GF.GR}), one has (suppressing the
  multi-indices to lighten the notation)
  \begin{subequations}
  \label{eq:greendefs}
    \begin{align}
      \GG_{\text{R}}&= \GG_{\text{F}}+i\WW_- = -\GG_{\text{D}}-i\WW_{+}, \\[3pt]
      \GG_{\text{A}}&= \GG_{\text{F}}+i\WW_{+} = -\GG_{\text{D}}-i\WW_{-}, \\[3pt]
        \GG_{\text{F}} &= \frac{1}{2} \left( \GG_{\text{R}} + \GG_{\text{A}} \right) - \frac{i}{2} \left( \WW_{+} + \WW_{-} \right), \\[3pt]
        \GG_{\text{F}} + \GG_{\text{D}} &= -i \left( \WW_{+} + \WW_{-} \right).
    \end{align}
\end{subequations}

\subsection{Flat spacetime}

The Wightman functions $\WW_\pm$ of a bulk field in flat spacetime can be
expressed in a form that makes manifest their large $r \equiv |\vb{r} - \vb{r}'|$
behavior. Suppressing once again the multi-indices, this expression is given by 
\begin{equation}\label{eq:Wmink}
    \begin{split}
        \WW_{\pm}(x,x') &= \int \frac{\dd[3]{\vb{k}}}{(2\pi)^3} \int \frac{\dd{\omega}}{2\pi} \, \theta(\pm \omega) \delta(\omega^2 - k^2) \, e^{-i\omega (t-t') + i\vb{k} \cdot (\vb{x} - \vb{x}')} \\[3pt]
        &= \frac{1}{2\pi r} \int_{0}^{\infty} \frac{\dd{\omega}}{2\pi} \, e^{\mp i\omega (t-t')} \sin(\omega r)\,, 
    \end{split}
\end{equation}
from which one can find
\begin{subequations}\label{eq:Gmink}
  \begin{align}
        \GG_{\text{R}}(x-x') &= - \frac{1}{4\pi r} \int_{\omega} e^{-i\omega(t-t'-r)}, \\[3pt]
        \GG_{\text{A}}(x-x') &= - \frac{1}{4\pi r} \int_{\omega} e^{-i\omega(t-t'+r)}, \\[3pt]
        \GG_{\text{F}}(x-x') &= - \frac{1}{4\pi r} \int_{\omega} e^{-i\omega(t-t')} e^{i|\omega|r}, \\[3pt]
        \GG_{\text{D}}(x-x') &= \frac{1}{4\pi r} \int_{\omega} e^{-i\omega(t-t')} e^{-i|\omega|r}.
    \end{align}
\end{subequations}

\subsection{One-dimensional field theory}

Let's specialize the previous formula -- in particular, eq.~\eqref{eq:W_pm.syms} -- to the theory of worldline operators, such as the multipole moments of a BH. Denoting the Wightman function of these operators by $\Delta$, one can write both $\Delta_{\pm}^{LL'}$ in terms of a single $\Delta^{LL'}$ function as
\begin{align}
  \Delta_{+}^{LL'}(\tau) & =\Delta^{LL'}(\tau)=\int\frac{\dd{\omega}}{2\pi} \, e^{-i\omega \tau}
  \tilde\Delta^{LL'}(\omega), \\
\Delta_{-}^{LL'}(\tau) & =\Delta^{L'\!L}(-\tau)=\int\frac{\dd{\omega}}{2\pi} \, e^{-i\omega \tau} \tilde\Delta^{L'\!L}(-\omega).
\end{align}
The dispersive representation of the retarded Green function is
\begin{align*}
G_{\text{R}}^{LL'}(\tau) &= -i\theta(\tau) \left( \Delta^{LL'}(\tau) - \Delta^{L'\!L}(-\tau) \right) \\[3pt]
                         &= \int\frac{\dd\omega}{2\pi}\frac{e^{-i\omega\tau}}{\omega+i\varepsilon} \left( \int\frac{\dd\omega'}{2\pi}\tilde{\Delta}^{LL'}(\omega')e^{-i\omega' \tau} -\int\frac{\dd\omega'}{2\pi}\tilde{\Delta}^{L'\!L}(-\omega') e^{-i\omega'\tau} \right) \\[3pt]
                         &= \int\frac{\dd\omega}{2\pi} \, e^{-i\omega\tau}\int\frac{\dd\omega'}{2\pi} \frac{\tilde{\Delta}^{LL'}(\omega')-\tilde{\Delta}^{L'\!L}(-\omega')}{\omega-\omega'+i\varepsilon}.
\end{align*}
It follows that
\begin{equation}\label{eq:GRdisp}
    \tilde{G}_{\text{R}}^{LL'}(\omega) = \int\frac{\dd\omega'}{2\pi} \frac{\tilde{\Delta}^{LL'}(\omega') - \tilde{\Delta}^{L'\!L}(-\omega')}{\omega-\omega'+i\varepsilon}.
\end{equation}

\subsection{Two-point functions in the in-in formalism}

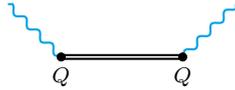
\begin{figure}
  \begin{center}
      \begin{tikzpicture}
      \begin{feynman}
        \draw[Cerulean,photon,thick](-0.8,0)--(-1.5,0.7);
        \draw[Cerulean,photon,thick](0.8,0)--(1.5,0.7);
        \draw[thick,double](-0.8,0)--(0.8,0);
        \filldraw[black] (-0.8,0) circle (1.5pt) node[anchor=south]{};
        \filldraw[black] (0.8,0) circle (1.5pt) node[anchor=south]{};
        \draw(-0.8,0)node[anchor=north]{\scriptsize $Q$};
        \draw(0.8,0)node[anchor=north]{\scriptsize $Q$};
      \end{feynman}
      \end{tikzpicture}
    \caption{Schematic representation of the GW scattering process mediated by
      two worldline insertions of the quadrupole operator.}
    \label{fig:int}
  \end{center}
\end{figure}

To correctly implement causal boundary conditions, one needs to write the two-point Wightman functions of a bulk field in the in-in formalism \cite{Keldysh:1964ud,Schwinger:1960qe}. Technically, this is implemented by doubling the degrees of freedom (d.o.f), so that the original d.o.f, labeled ``1'', propagates forward in time, while the additional d.o.f labeled ``2'', propagate backwards in time. In the $\{a,b\} = \{1,2\}$ basis the two-point functions for the worldline fields are given by the matrix (dropping any Lorentz index)
\begin{equation}
    G^{ab} =
    \begin{pmatrix}
        G_\text{F} & i\Delta_{-} \\
        i\Delta_{+} & G_\text{D}
    \end{pmatrix},
\end{equation}
which in the Keldysh representation\footnote{This basis is defined by $x_{+} \equiv \frac{1}{2}(x_1 + x_2)$ and $x_{-} \equiv x_1 - x_2$.} $\{A,B\} = \{+,-\}$ becomes
\begin{equation}
    G^{AB} = 
    \begin{pmatrix}
        0 & G_{\text{A}} \\
        G_{\text{R}} & -i(\Delta_{+} + \Delta_{-})
    \end{pmatrix}.
\end{equation}
In this basis, the in-in indices are raised and lowered with the anti-diagonal tensor $\epsilon_{AB}$, so that  
\begin{equation}
    G_{AB} \equiv \epsilon_{AC} G^{CD} \epsilon_{DB} = 
    \begin{pmatrix}
        -i(\Delta_{+} + \Delta_{-}) & G_{\text{R}} \\
        G_{\text{A}} & 0
    \end{pmatrix}
    , \quad 
    \epsilon_{AB} \equiv 
    \begin{pmatrix}
        0 & 1 \\ 
        1 & 0
    \end{pmatrix}.
\end{equation}
Finally one can write
\begin{equation}
    G_{ab} = 
    \begin{pmatrix}
        G_{\text{F}} & -i\Delta_{-} \\ 
        -i\Delta_{+} & G_{\text{D}}
    \end{pmatrix},
\end{equation}
with analog formulae for the bulk Green functions $\GG$. The bulk field two-point function $\WW_{\text{BH}}$, corrected for the leading-order interaction with the source\footnote{In order to recover the functions $G_{ab}$ of \cite{Goldberger:2019sya} one must multiply our functions $G_{ab}$ and $\GG_{ab}$ by a factor of $i$.}
as per the process in fig.~\ref{fig:int} can be written perturbatively in terms of their Minkowskian counterpart $\WW_{\text{M}}$ as
\begin{equation}\label{eq:2ptinin}
    \begin{split}
        -i\WW^{AB}_{\text{BH}}(x,x') &= -i\WW^{AB}_{\text{M}}(x,x') \\
        &- i\int \dd[4]{\tau} \dd[4]{\tau'} \GG^{AC}(t,\vb{ x},\tau,0) G_{CD}(\tau,\tau') \GG^{DB}(\tau',0,t',\vb{x'}). 
    \end{split}
\end{equation}
Repeated in-in indices are summed over and we assume that the BH is sitting at $\vb{x} = 0$. Denote $|\vb{x}| = r$ and consider $A = 1$ and $B=2$. Then, in the limit $r = r' \to \infty$ the right-hand side above becomes  
\begin{equation}
    \WW^{12}_M(t-t',0) + \frac 1{(4\pi r)^2} \int \frac{\dd{\omega}}{2\pi} \, e^{-i\omega(t-t')} \left[ \theta(-\omega) \Delta_{+}(\omega) + \theta(\omega)\Delta_{-}(\omega) \right],
\end{equation}
as expressed in eq. \eqref{eq:bulk2pt}. To derive this formula the expressions for $\GG_{ab},\WW_{ab},G_{ab}$ in terms of $\WW_\pm,\Delta_\pm$ have been used, as well as the expression \eqref{eq:Wmink} and \eqref{eq:Gmink} for the Minkowski two-point function of the bulk field in the aforementioned limit. 

\providecommand{\href}[2]{#2}\begingroup\raggedright\endgroup


\end{document}